\documentclass{article}
\usepackage[T1]{fontenc}
\UseRawInputEncoding
\usepackage[english]{babel}
\usepackage{amsmath}
\usepackage{amsthm}
\usepackage[square,numbers]{natbib}
\makeatletter
\numberwithin{equation}{section}
\numberwithin{figure}{section}

\usepackage{url}
\usepackage{hyperref}

\makeatother

\begin{document}
\begin{flushleft}
\textbf{OUJ-FTC-16}\\
\ \\
 
\par\end{flushleft}

\begin{center}
{\LARGE{}A Lorentz Covariant Matrix Model for Bosonic M2-Branes:
Nambu Brackets and Restricted Volume-Preserving Deformations}{\LARGE\par}
\par\end{center}

\begin{center}
\vspace{10pt}
\par\end{center}

\begin{center}
So Katagiri\textsuperscript{}\footnote{So.Katagiri@gmail.com}
\par\end{center}

\begin{center}
\textit{$^{*}$Nature and Environment, Faculty of Liberal Arts, The
Open University of Japan, Chiba 261-8586, Japan} \\
 
\par\end{center}

\vspace{10pt}

\begin{abstract}
We propose a Lorentz covariant matrix model as a nonperturbative formulation
of the bosonic M2-brane in M-theory. Unlike previous approaches relying
on the light-cone gauge or symmetry-based constructions, our model
retains full 11-dimensional Lorentz invariance by introducing a novel
gauge-fixing condition that restricts the symmetry of volume-preserving
deformations (VPD) to a subclass, which we call restricted VPD (RVPD).
This restriction enables a consistent matrix regularization of the
Nambu bracket, bypassing the long-standing obstructions related to
the Leibniz rule and the Fundamental Identity. The resulting model
exhibits RVPD symmetry, admits particle-like and noncommutative membrane
solutions, and lays the foundation for a Lorentz-invariant, nonperturbative
matrix description of M2-branes.

Our work offers a new paradigm for constructing Lorentz-invariant
matrix models of membranes, revisiting the algebraic structure underlying
M-theory.
\end{abstract}

\section{Introduction}

In modern particle physics, the fundamental particles that make up
the microscopic world are quarks and leptons, which interact with
each other through gauge particles. One of the significant features
of string theory is that these particles can be understood as different
excitation states of ``strings,'' which are objects extended in one
dimension. Various types of string theories are known, including Type
I, Type IIA, Type IIB, and Heterotic string theories.

To understand why such different types of string theories exist, there
is an approach that considers ``membranes,'' which are objects extended
in two dimensions, and aims to interpret the various types of string
theories as different excitation states of these membranes. This promising
approach is called ``M-theory.'' The ``M'' in M-theory is said to
stand for ``Membrane'' or ``Mother'' (as in ``Mother theory'').

The successful formulation of this M-theory was achieved by the BFSS
model. ``BFSS'' is named after the initials of the four researchers,
including Banks, who developed the model \citep{Banks_1997} .

This M-theory was formulated using the so-called light cone gauge,
breaking the spacetime symmetry. However, if it were possible to formulate
the theory without breaking spacetime symmetry, it would provide a
clearer understanding of the structure of spacetime, which is desirable.
This is referred to as a \textquotedbl Lorentz covariant formulation
of M-theory.\textquotedbl{} Unfortunately, such a formulation has
not yet been achieved.

The BFSS model can be viewed as a matrix regularized theory of membranes
under the light cone gauge \citep{de_Wit_1988}. However, since taking
the light cone gauge breaks Lorentz covariance, obtaining a Lorentz
covariant non-perturbative model has remained a long-standing problem\footnote{There have been previous discussions on Lorentz covariant matrix models
\citep{Fujikawa_1997,Awata_1998,Minic2000,Smolin_2000,Yoneya_2016,Ashwinkumar_2021},
but the field remains in an exploratory stage.}.

It has been known that M2-branes can be rewritten using the Nambu
bracket. There has long been hope that a Lorentz covariant matrix
model could be obtained by \textquotedbl quantizing\textquotedbl{}
(matrix regularizing) the Nambu bracket \citep{Minic2000,Awata_2001}.
However, it is also known that it is challenging to quantize the Nambu
bracket while preserving its inherent complete antisymmetry, the Leibniz
rule, and the Fundamental Identity. This issue remains an unresolved
problem to this day. As a result, the path to constructing a matrix
model through the quantization of the Nambu bracket has remained closed\footnote{For example, in 1973, Nambu initially proposed a ternary relation
as a quantum Nambu bracket, later known as the Heisenberg-Nambu bracket.
However, he reported the breaking of the Leibniz rule and also examined
the non-associativity of the algebra in his paper \citep{Nambu_1973}.
It was Takhtajan who pointed out that, in addition to complete antisymmetry
and the Leibniz rule, the Fundamental Identity is also required for
the Nambu bracket \citep{Takhtajan_1994}. They proposed a Zariski
quantization using Zariski algebras and suggested a quantum Nambu
bracket that satisfies these properties, but its physical meaning
remains unclear \citep{Dito_1997}.

Other approaches include studies using cubic matrices \citep{Awata_2001},
analyses with the Hamilton-Jacobi formalism \citep{Yoneya_2017},
and investigations using path integrals \citep{Saitou_2014}. Around
2008, the BLG model was proposed, introducing a theory using Lie 3-algebras
as the low-energy effective theory of M2-branes. Since its infinite-dimensional
representation becomes the Nambu bracket, the relationship between
M2-branes and the Nambu bracket attracted attention \citep{Bagger_2007,Gustavsson_2009}.

However, the BLG model is a low-energy effective theory for multiple
M2-branes derived from symmetry considerations, and its relationship
with the matrix model as a non-perturbative theory of M-theory remains
unresolved. The BLG model is believed to describe only two M2-branes
due to arguments based on group structure. Subsequently, the ABJM
model proposed a description of N M2-branes as a Chern-Simons model
with a bifundamental gauge group, but its connection with Lie 3-algebras
and the Nambu bracket has become somewhat distant \citep{Aharony_2008}.

More recently, an operator-based formulation of the Nambu bracket
quantization within classical mechanics was proposed, offering a novel
perspective on the canonical structure underlying Nambu dynamics\citep{Katagiri2022}.}.

In this study, we avoid this problem by obtaining a matrix model through
partially constraining the volume-preserving deformation (RVPD) of
the membrane action using gauge-fixing conditions\footnote{In this study, the quantization of the Nambu bracket refers to the
matrix regularization of the membrane action, and the quantization
of the membrane itself requires further investigation.}.

The obtained matrix model preserves 11-dimensional Lorentz invariance
and possesses the symmetry (RVPD) arising from the constrained volume-preserving
deformation. The solutions to its equations of motion include configurations
with a two-dimensional non-commutative extension.

This study has two key features:
\begin{enumerate}
\item The long-standing issues in the quantum Nambu bracket, such as the
violation of the Leibniz rule and the Fundamental Identity (F.I.),
are shown to be not essential for the consistency of the theory when
only the canonical quantization procedure is used.
\item Even in the action of such membranes, the most critical symmetry of
the membrane, the volume-preserving deformation, is maintained in
a restricted form (RVPD), and a Lorentz-invariant matrix model for
membranes is achieved.
\end{enumerate}
This research offers insights into a non-perturbative definition of
M-theory and aims to provide a starting point for obtaining a Lorentz-invariant
matrix model for membranes. The exploration of connections with other
matrix models and extensions to supersymmetry remain as future challenges.

The structure of this study is as follows:
\begin{itemize}
\item Section 2: The action of the membrane is described using the Nambu
bracket, and the Nambu bracket is decomposed using the Poisson bracket.
\item Section 3: Gauge-fixing conditions are introduced to further restrict
the volume-preserving deformation (VPD), resulting in a restricted
volume-preserving deformation (RVPD).
\item Section 4: The properties of the RVPD are discussed, demonstrating
that it satisfies the composition rule of transformations. A matrix
regularization is performed to obtain the matrix model.
\item Section 5: The equations of motion are derived, and solutions such
as particle solutions and non-commutative membranes are examined.
\item Section 6: The conclusions are summarized.
\item Section 7: Further discussions are presented.
\item Appendix A: The algebraic aspects of RVPD are analyzed.
\item Appendix B: The necessity and sufficiency of the gauge-fixing condition
for restricting the gauge parameters are proven.
\end{itemize}

\section{The action of the membrane}

A two-dimensional membrane in an 11-dimensional spacetime can be expressed
using the Nambu bracket by appropriately gauge-fixing the bosonic
part of the Nambu-Goto-type membrane action as follows:

\begin{equation}
S=\int d^{3}\sigma\frac{1}{2}\{X^{I},X^{J},X^{K}\}^{2}.
\end{equation}
Here, $X^{I}(\sigma^{1},\sigma^{2},\sigma^{3})$ represents the spacetime
coordinates of the membrane, where $I=0,\dots,10$. The parameters
$\sigma^{i}$ with $i=1,2,3$ are the internal parameters of the membrane.
The expression $\{X^{I},X^{J},X^{K}\}$ is called the Nambu bracket
and is defined by

\begin{equation}
\{X^{I},X^{J},X^{K}\}=\epsilon^{ijk}\frac{\partial X^{I}}{\partial\sigma^{i}}\frac{\partial X^{J}}{\partial\sigma^{j}}\frac{\partial X^{K}}{\partial\sigma^{k}}.
\end{equation}
This action is invariant under the volume-preserving deformation (VPD),
represented by the transformation:
\begin{equation}
\delta_{\mathrm{VPD}}X^{I}=\{Q_{1},Q_{2},X^{I}\}
\end{equation}
where $Q_{1}$ and $Q_{2}$ are arbitrary charges.

To perform matrix regularization, we want to decompose this Nambu
bracket using the Poisson bracket. Here, the Poisson bracket for the
two components $\sigma^{1}$ and $\sigma^{2}$ among the three components
is defined as:

\begin{equation}
\{A,B\}=\epsilon^{ab}\frac{\partial A}{\partial\sigma^{a}}\frac{\partial B}{\partial\sigma^{b}}
\end{equation}
where $a=1,2$. 

However, a straightforward decomposition of the Nambu bracket using
the Poisson bracket followed by matrix regularization causes a loss
of the Fundamental Identity (F.I.) that the Nambu bracket originally
possessed. This leads to a breakdown of the transformation properties
related to the composition of deformations.

In this study, we avoid this problem by rewriting the Nambu bracket
into a special form using the Poisson bracket and then partially restricting
the volume-preserving deformation. This approach preserves the essential
properties of the transformation and allows for a consistent matrix
model.

Specifically, the Nambu bracket can be rewritten as follows:\footnote{While this decomposition is motivated by the structure of Takhtajan's
action, which corresponds to the Hamiltonian formulation of Nambu
mechanics, it is important to note that the decomposition itself can
be regarded as a purely algebraic transformation, independent of any
specific dynamical framework.

Takhtajan's action serves as the Hamiltonian formulation of Nambu
mechanics and has been studied since the era of Nambu and Sugamoto
\citep{Nambu_1980,Sugamoto_1983,Takhtajan_1994}. The relationship
between Takhtajan's action and the membrane action is analogous to
that between the Hamiltonian formulation of classical mechanics and
the Schild action in string theory. Just as the quantization of the
Poisson bracket is naturally considered in the Hamiltonian framework
when examining the matrix regularization of strings, it is also meaningful
to first consider the quantization of the Nambu bracket within Takhtajan's
action before proceeding to the matrix regularization of the membrane.

Regarding the quantization of Takhtajan's action, Sakakibara discussed
it from the perspective of deformation quantization \citep{Sakakibara_2000},
and later, Matsuo and Shibusa explored canonical quantization in the
$x^{3}=\sigma$ gauge \citep{Matsuo_2001}.}
\begin{equation}
\{X^{I},X^{J},X^{K}\}=\{\tau(X^{I},X^{J}),X^{K}\}+\frac{\partial X^{K}}{\partial\sigma^{3}}\{X^{I},X^{J}\}+\Sigma(X^{I},X^{J};X^{K})
\end{equation}
where
\begin{equation}
\Sigma(A,B;C)\equiv A\{\frac{\partial B}{\partial\sigma^{3}},C\}-B\{\frac{\partial A}{\partial\sigma^{3}},C\}
\end{equation}
and

\begin{equation}
\tau(A,B)\equiv\frac{\partial A}{\partial\sigma^{3}}B-\frac{\partial B}{\partial\sigma^{3}}A.
\end{equation}
Thus, the volume-preserving deformation (VPD) takes the form:

\begin{equation}
\delta_{\mathrm{VPD}}X^{I}\equiv\{Q_{1},Q_{2},X^{I}\}=\{\tau(Q_{1},Q_{2}),X^{I}\}+\frac{\partial X^{I}}{\partial\sigma^{3}}\{Q_{1},Q_{2}\}+\Sigma(Q_{1},Q_{2};X^{I}).
\end{equation}

As is evident from this reformulation using the Poisson bracket, if
this is directly applied to the canonical commutation relations, the
composition rule of transformations would be violated. This reflects
the fact that the Fundamental Identity (F.I.) of the Nambu bracket
is broken under quantization.

If the volume-preserving deformation (VPD) is restricted to a certain
subclass that operates under the Poisson bracket, which we refer to
as the restricted volume-preserving deformation (RVPD), then it becomes
possible to achieve a Lorentz-covariant quantization of the membrane
while preserving Lorentz covariance. In the following sections, we
demonstrate this explicitly

\section{Gauge Fixing Condition and Restriction of Gauge Parameters}

In this study, we further gauge-fix the volume-preserving deformation
(VPD) to ensure that, even when the deformation is incorporated into
the canonical commutation relations as a restricted deformation, the
composition rule remains intact. Our goal is to retain only those
deformations with well-behaved properties.

We refer to such a restricted volume-preserving deformation as \textit{RVPD
(Restricted Volume-Preserving Deformation)} and denote it by $\delta_{R}X^{I}$.
When specifying the associated charges explicitly, we write this as
$\delta_{R(Q_{1},Q_{2})}X^{I}$.

The first essential property of a well-behaved deformation is \textit{linearity,}
expressed as:
\begin{equation}
\delta_{R}(\lambda_{1}A_{1}+\lambda_{2}A_{2}+\dots)=\lambda_{1}\delta_{R}A_{1}+\lambda_{2}\delta_{R}A_{2}+\dots
\end{equation}
Next, it is natural to require that the deformation is \textit{distributive},
satisfying:

\begin{equation}
\delta_{R}\left(A_{1}A_{2}\dots A_{n}\right)=\left(\delta_{R}A_{1}\right)A_{2}\dots A_{n}+A_{1}\left(\delta_{R}A_{2}\right)\dots A_{n}+\dots+A_{1}A_{2}\dots\left(\delta_{R}A_{n}\right).
\end{equation}
These properties are consistent with the distributive law of the classical
Nambu bracket.

Additionally, the deformation should preserve the properties of the
classical Nambu bracket that arise from its complete antisymmetry,
including:
\begin{itemize}
\item Charge preservation:
\end{itemize}
\begin{equation}
\delta_{R(Q_{1},Q_{2})}Q_{1,2}=0
\end{equation}

\begin{itemize}
\item Charge exchange symmetry:
\begin{equation}
\delta_{R(Q_{1},Q_{2})}X^{I}=-\delta_{R(Q_{2},Q_{1})}X^{I}
\end{equation}
\end{itemize}
Finally, the most important property is the composition rule of transformations,
which ensures that the Fundamental Identity (F.I.) of the classical
Nambu bracket is maintained even after repeated deformations. This
rule plays a role analogous to the Jacobi identity in matrix algebras:
\begin{equation}
\{Q,\{H,X\}\}=\{\{Q,H\},X\}+\{H,\{Q,X\}\}.
\end{equation}
The composition rule for the RVPD is expressed as:

\begin{equation}
\begin{aligned}\delta_{R(Q_{1},Q_{2})} & \delta_{R(H_{1},H_{2})}X^{I}=\\
 & \delta_{R(\delta_{R(Q_{1},Q_{2})}H_{1},H_{2})}X^{I}+\delta_{R(H_{1},\delta_{R(Q_{1},Q_{2})}H_{2})}X^{I}+\delta_{R(H_{1},H_{2})}\delta_{R(Q_{1},Q_{2})}X^{I}.
\end{aligned}
\end{equation}

This property ensures that the composition of transformations under
the RVPD remains consistent and maintains the algebraic structure
of the theory.

When the volume-preserving deformation (VPD) is straightforwardly
applied through matrix regularization, it takes the form:
\begin{equation}
\delta_{\mathrm{VPD}}X^{I}=[\tau(Q_{1},Q_{2}),X^{I}]+\frac{\partial X^{I}}{\partial\sigma^{3}}[Q_{1},Q_{2}]+\Sigma(Q_{1},Q_{2};X^{I}).
\end{equation}

It is evident that this formulation violates all of the desired properties
except for charge exchange symmetry.

Among the terms, the component that best preserves the distributive
property of the deformation is:
\begin{equation}
[\tau(Q_{1},Q_{2}),X^{I}].
\end{equation}
Therefore, we aim to perform gauge fixing in such a way that this
term is retained while the problematic terms are eliminated.

However, the term introduces an\textbf{ }\textit{incomplete Leibniz
rule} between the $\tau$ bracket and the commutator $[,]$, as shown
below:
\begin{equation}
[\tau(A,B),C]=\tau([A,C],B)+\tau(A,[B,C])+\Delta(A,B;C).
\end{equation}
Here, $\Delta$ quantifies the violation of the Leibniz rule and is
defined by:
\begin{equation}
\Delta(A,B;C)\equiv A[B,\frac{\partial C}{\partial\sigma^{3}}]-B[A,\frac{\partial C}{\partial\sigma^{3}}] - [[A,B],\frac{\partial C}{\partial\sigma^{3}}] .
\end{equation}
A similar discussion is presented in Sakakibara's work \citep{Sakakibara_2000}.
In the present paper, it is found that the discrepancy $\Delta(A,B;C)$
vanishes when $C$ is independent of $\sigma^{3}$.

This behavior impacts the composition of transformations, leading
to the following relation:
\begin{equation}
\begin{aligned}[][\tau(Q_{1},Q_{2}), & \tau(H_{1},H_{2})]=\\
 & \tau([\tau(Q_{1},Q_{2}),H_{1}],H_{2})+\tau(H_{1},[\tau(Q_{1},Q_{2}),H_{2}])+\Delta(H_{1},H_{2};\tau(Q_{1},Q_{2})).
\end{aligned}
\end{equation}

The deviation caused by the incomplete Leibniz rule is proportional
to $\Delta$.

To eliminate this discrepancy, we must impose a restriction on the
gauge parameters such that: 
\begin{equation}
\frac{\partial\tau(Q_{1},Q_{2})}{\partial\sigma^{3}}=0.
\end{equation}
This condition effectively removes the $\sigma^{3}$ dependence from
the volume-preserving deformation, ensuring the consistency of the
transformation composition.

Focusing on the term $\{\tau(Q_{1},Q_{2}),X^{I}\}$ implies that the
remaining terms must also satisfy a gauge parameter restriction such
that:
\begin{equation}
\frac{\partial X^{I}}{\partial\sigma^{3}}\{Q_{1},Q_{2}\}+\Sigma(Q_{1},Q_{2};X^{I})=0.
\end{equation}
Additionally, considering the condition for \textit{charge preservation}:

\begin{equation}
\delta_{R(Q_{1},Q_{2})}Q_{1,2}=0,
\end{equation}
it becomes clear that the previous conditions must individually hold
as:
\begin{equation}
\{Q_{1},Q_{2}\}=0,
\end{equation}

\begin{equation}
\Sigma(Q_{1},Q_{2};X^{I})=0.
\end{equation}

Consequently, the problem transforms into an inverse problem, where
we seek an appropriate gauge-fixing condition that can impose the
following restrictions on the gauge parameters:
\begin{equation}
\frac{\partial\tau(Q_{1},Q_{2})}{\partial\sigma^{3}}=0,
\end{equation}

\begin{equation}
\{Q_{1},Q_{2}\}=0,
\end{equation}

\begin{equation}
\Sigma(Q_{1},Q_{2};X^{I})=0.
\end{equation}

Since there are two charges involved, it is evident that only one
gauge-fixing condition is needed. Moreover, this condition must preserve
Lorentz invariance. Given that the restriction $\frac{\partial\tau(Q_{1},Q_{2})}{\partial\sigma^{3}}=0$
eliminates $\sigma_{3}$ dependence in $\tau(Q_{1},Q_{2})$, it is
natural to consider a gauge condition that constrains the motion of
$X^{I}$ in the $\sigma^{3}$ direction.

The simplest gauge-fixing condition can be formulated using a constant
$C_{I}$ as:\footnote{The gauge fixing using a fixed vector $C_{I}$ is formally analogous
to the axial gauge condition $n^{\mu}A_{\mu}=0$, which employs a
fixed vector $n^{\mu}$. However, in the axial gauge, $n^{\mu}$ is
treated as a fixed background and is not transformed under Lorentz
transformations, which breaks the Lorentz covariance of the theory. 

In contrast, the present work treats $C_{I}$ consistently as a Lorentz
vector, ensuring that the gauge-fixed action remains Lorentz covariant.
Therefore, the role of the fixed vector in this study is fundamentally
different from that in the conventional axial gauge.}\footnote{The introduction of the constant vector $C_I$ in the gauge restriction does not break Lorentz covariance. 
Unlike the fixed $n_\mu$ in the axial gauge, here $C_I$ is treated as a genuine Lorentz vector and consistently transforms under Lorentz transformations. 
The gauge-fixed action therefore remains Lorentz covariant. 
This is not a naive assumption: the situation is analogous to the light-cone gauge, where manifest covariance is hidden but the full physical Lorentz invariance is preserved. 
Thus, there is no inconsistency in adopting such a condition within the RVPD framework.}
\begin{equation}
C_{I}X^{I}=\sigma^{3}.
\end{equation}
Under this condition, the relation $C_{I}\delta_{R}X^{I}=0$ leads
to the requirement: 

\begin{equation}
\{Q_{1},Q_{2}\}=0.
\end{equation}
However, this condition alone does not provide the necessary additional
restrictions and to limit the $\sigma^{3}$ dependence of $\tau(Q_{1},Q_{2})$
and to eliminate the $\Sigma$ term, it is more effective to impose
a constraint not directly on $X^{I}$ but rather on its derivative
in the $\sigma^{3}$ direction:
\begin{equation}
C_{I}\partial_{\sigma^{3}}X^{I}=\sigma^{3}.
\end{equation}

Physically, this gauge-fixing condition implies that the membrane's
motion along the direction defined by the Lorentz vector $C_{I}$,
with $\sigma^{3}$ regarded as a temporal evolution parameter, corresponds
to uniformly accelerated motion. This contrasts with the BFSS model,
where gauge degrees of freedom are reduced by adopting a light-cone
frame moving at the speed of light. In the present work, by choosing
a uniformly accelerated frame instead, we achieve a consistent restriction
of the gauge parameters associated with volume-preserving deformations\footnote{This interpretation is due to a remark by Associate Professor Shiro
Komata, to whom I am deeply grateful.}.

With this gauge-fixing condition, $C_{I}X^{I}$ can be expressed using
an appropriate function $f(\sigma^{1},\sigma^{2})$ as:

\begin{equation}
C_{I}X^{I}=\frac{1}{2}\left(\sigma^{3}\right)^{2}+f(\sigma^{1},\sigma^{2}).
\end{equation}
By analyzing the independence of $f(\sigma^{1},\sigma^{2})$ and considering
the independence of each term, we can derive the following restrictions
on the gauge parameters $Q_{1},Q_{2}$:
\begin{equation}
\partial_{\sigma^{3}}\tau(Q_{1},Q_{2})=0,
\end{equation}

\begin{equation}
\partial_{a}\partial_{\sigma^{3}}Q_{1,2}=0,\ a=1,2.
\end{equation}

These constraints ensure that the deformation retains the desired
properties while maintaining the composition rule of transformations
and preserving Lorentz invariance.

This gauge-fixing condition allows us to restrict the volume-preserving
deformation in such a way that the composition rule of transformations
is maintained as much as possible, thereby approximating the Fundamental
Identity (F.I.) of the Nambu bracket even after quantization.

When the volume-preserving deformation (VPD) is applied under the
gauge-fixing condition, it results in the following expression:

\begin{equation}
\begin{aligned}C_{I}\partial_{\sigma^{3}} & \delta_{\mathrm{VPD}}X^{I}=\partial_{\sigma^{3}}\left(\{\tau(Q_{1},Q_{2}),C_{I}X^{I}\}\right)\\
 & +\partial_{\sigma^{3}}\left(C_{I}\frac{\partial X^{I}}{\partial\sigma^{3}}\{Q_{1},Q_{2}\}\right)+\partial_{\sigma^{3}}\left(\Sigma(Q_{1},Q_{2};C_{I}X^{I})\right)
\end{aligned}
\end{equation}
Expanding this expression gives:

\begin{equation}
\begin{aligned}= & \{\partial_{\sigma^{3}}\tau(Q_{1},Q_{2}),C_{I}X^{I}\}+\{\tau(Q_{1},Q_{2}),C_{I}\partial_{\sigma^{3}}X^{I}\}\\
 & +\{Q_{1},Q_{2}\}+\sigma^{3}\partial_{\sigma^{3}}\{Q_{1},Q_{2}\}\\
 & +\Sigma(\partial_{\sigma^{3}}Q_{1},Q_{2};C_{I}X^{I})+\Sigma(Q_{1},\partial_{\sigma^{3}}Q_{2};C_{I}X^{I})+\Sigma(Q_{1},Q_{2};C_{I}\partial_{\sigma^{3}}X^{I}).
\end{aligned}
\end{equation}

Since we have the conditions:

\begin{equation}
\{\tau(Q_{1},Q_{2}),C_{I}\partial_{\sigma^{3}}X^{I}\}=0,\ \Sigma(Q_{1},Q_{2};C_{I}\partial_{\sigma^{3}}X^{I})=0,
\end{equation}
 the expression can be simplified to:

\begin{equation}
\begin{aligned}\{\partial_{\sigma^{3}}\tau(Q_{1},Q_{2}),C_{I}X^{I}\} & +\{Q_{1},Q_{2}\}+\sigma^{3}\partial_{\sigma^{3}}\{Q_{1},Q_{2}\}\\
 & +\Sigma(\partial_{\sigma^{3}}Q_{1},Q_{2};C_{I}X^{I})+\Sigma(Q_{1},\partial_{\sigma^{3}}Q_{2};C_{I}X^{I})=0.
\end{aligned}
\end{equation}
Expanding $C_{I}X^{I}$ within the Poisson bracket, we obtain:

\begin{equation}
\begin{aligned}\{\partial_{\sigma^{3}}\tau(Q_{1},Q_{2}) & ,f(\sigma^{1},\sigma^{2})\}+\{Q_{1},Q_{2}\}+\sigma^{3}\partial_{\sigma^{3}}\{Q_{1},Q_{2}\}\\
 & +\Sigma(\partial_{\sigma^{3}}Q_{1},Q_{2};f(\sigma^{1},\sigma^{2}))+\Sigma(Q_{1},\partial_{\sigma^{3}}Q_{2};f(\sigma^{1},\sigma^{2}))=0.
\end{aligned}
\end{equation}
This implies that both the coefficients of $f(\sigma^{1},\sigma^{2})$
and the other terms must independently be zero:

\begin{equation}
\begin{aligned}\{\partial_{\sigma^{3}}\tau(Q_{1},Q_{2}),f(\sigma^{1},\sigma^{2})\} & +\Sigma(\partial_{\sigma^{3}}Q_{1},Q_{2};f(\sigma^{1},\sigma^{2})\}\\
 & +\Sigma(Q_{1},\partial_{\sigma^{3}}Q_{2};f(\sigma^{1},\sigma^{2}))=0
\end{aligned}
\end{equation}

\begin{equation}
\{Q_{1},Q_{2}\}+\sigma^{3}\partial_{\sigma^{3}}\{Q_{1},Q_{2}\}=0.
\end{equation}

Next, considering the first equation, decomposing the Poisson bracket,
we obtain:

\begin{equation}
K_{(\tau)}^{b}\partial_{b}f+K_{\Sigma1}^{b}\partial_{b}f+K_{\Sigma2}^{b}\partial_{b}f=0,\label{eq:threeConstraintConditonSum}
\end{equation}
where the coefficients of $\partial_{b}f$ are constructed from the
following differential expressions involving the gauge parameters:

\begin{equation}
K_{(\tau)}^{b}\equiv\epsilon^{ab}\partial_{a}\partial_{\sigma^{3}}\tau(Q_{1},Q_{2}),
\end{equation}

\begin{equation}
K_{\Sigma1}^{b}\equiv\partial_{\sigma^{3}}Q_{1}\epsilon^{ab}\partial_{a}\partial_{\sigma^{3}}Q_{2}-Q_{2}\epsilon^{ab}\partial_{a}\partial_{\sigma^{3}}^{2}Q_{1},
\end{equation}

\begin{equation}
K_{\Sigma2}^{b}\equiv Q_{1}\epsilon^{ab}\partial_{a}\partial_{\sigma^{3}}^{2}Q_{2}-\partial_{\sigma^{3}}Q_{2}\epsilon^{ab}\partial_{a}\partial_{\sigma^{3}}Q_{1}.
\end{equation}
Since the differentiation patterns in each term are distinct, they
can be considered formally independent. Thus, we conclude:
\begin{equation}
\{\partial_{\sigma^{3}}\tau(Q_{1},Q_{2}),f\}=0,
\end{equation}

\begin{equation}
\Sigma(\partial_{\sigma^{3}}Q_{1},Q_{2};f)=0,
\end{equation}

\begin{equation}
\Sigma(Q_{1},\partial_{\sigma^{3}}Q_{2};f)=0.
\end{equation}
These conditions provide the necessary constraints to preserve the
composition rule of the transformations even after the volume-preserving
deformation is restricted.

See Appendix B for a more mathematical discussion of the necessary
and sufficient conditions for the restriction of gauge parameters
from gauge constraints.

\section{Restricted Volume-Preserving Deformation (RVPD)}

With the restrictions on $Q_{1}$ and $Q_{2}$ established in the
previous section, we denote the restricted parameters as $Q_{1}^{(R)}$
and $Q_{2}^{(R)}$ . Under these restrictions, the Restricted Volume-Preserving
Deformation (RVPD) can be expressed as:
\begin{equation}
\delta_{R}X=\{Q_{1}^{(R)},Q_{2}^{(R)},X^{I}\}=\{\tau(Q_{1}^{(R)},Q_{2}^{(R)}),X^{I}\}.
\end{equation}
This formulation demonstrates that the deformation can be fully described
using only the term $\tau(Q_{1}^{(R)},Q_{2}^{(R)})$. As a result,
the original complexity of the Nambu bracket is significantly reduced,
while the transformation continues to maintain the desired properties,
such as the composition rule, under the restricted conditions.

Under this gauge-fixing condition, the action can be written as:

\begin{equation}
S=\int d^{3}\sigma\frac{1}{2}\left(\{\tau(X^{I},X^{J}),X^{K}\}+\frac{\partial X^{I}}{\partial\sigma^{3}}\{X^{J},X^{K}\}+\Sigma(X^{I},X^{J};X^{K})\right)^{2}
\end{equation}
with the gauge-fixing condition:

\begin{equation}
C_{I}\partial_{\sigma^{3}}X^{I}=\sigma^{3}.
\end{equation}
This action exhibits the symmetry of the Restricted Volume-Preserving
Deformation (RVPD) with:
\begin{equation}
\delta_{R}X^{I}=\{\tau(Q_{1}^{(R)},Q_{2}^{(R)}),X^{I}\},
\end{equation}
as well as global Lorentz invariance. The RVPD symmetry ensures that
the restricted deformation maintains the composition rule, while the
global Lorentz symmetry preserves the full 11-dimensional spacetime
invariance.

The following relation holds:

\begin{equation}
\begin{aligned}\{\tau(Q_{1},Q_{2}),\tau(H_{1},H_{2})\} & =\tau(\{\tau(Q_{1},Q_{2}),H_{1}\},H_{2})+\tau(H_{1},\{\tau(Q_{1},Q_{2}),H_{2}\})\\
 & +\Delta(H_{1},H_{2};\tau(Q_{1},Q_{2})).
\end{aligned}
\end{equation}
Since 
\begin{equation}
\frac{\partial}{\partial\sigma^{3}}\tau(Q_{1},Q_{2})=0,
\end{equation}
 the correction term $\Delta$ vanishes.

The composition of transformations under the RVPD is given by:

\begin{equation}
\delta_{Q^{(R)}}\delta_{H^{(R)}}X=\{\tau(Q_{1}^{(R)},Q_{2}^{(R)}),\{\tau(H_{1}^{(R)},H_{2}^{(R)}),X^{I}\}\}.
\end{equation}
Expanding this using the properties of the Poisson bracket:

\begin{equation}
=\{\{\tau(Q_{1}^{(R)},Q_{2}^{(R)}),\tau(H_{1}^{(R)},H_{2}^{(R)})\},X^{I}\}+\{\tau(H_{1}^{(R)},H_{2}^{(R)}),\{\tau(Q_{1}^{(R)},Q_{2}^{(R)}),X^{I}\}\}.
\end{equation}
Applying the incomplete Leibniz rule, we obtain:

\begin{equation}
\begin{aligned}= & \{\tau(\{\tau(Q_{1}^{(R)},Q_{2}^{(R)}),H_{1}^{(R)}\},H_{2}^{(R)}),X^{I}\}+\{\tau(H_{1},\{\tau(Q_{1}^{(R)},Q_{2}^{(R)}),H_{2}\}),X^{I}\}\\
 & +\{\tau(H_{1}^{(R)},H_{2}^{(R)}),\{\tau(Q_{1}^{(R)},Q_{2}^{(R)}),X^{I}\}\}.
\end{aligned}
\end{equation}
Rewriting this in terms of the RVPD transformations:

\begin{equation}
\begin{aligned}= & \{\tau(\delta_{Q^{(R)}}H_{1}^{(R)},H_{2}^{(R)}),X^{I}\}+\{\tau(H_{1}^{(R)},\delta_{Q^{(R)}}H_{2}^{(R)}),X^{I}\}\\
 & +\{\tau(H_{1}^{(R)},H_{2}^{(R)}),\delta_{Q^{(R)}}X^{I}\}.
\end{aligned}
\end{equation}

This demonstrates that the Leibniz rule is preserved under RVPD, and
the composition rule of the transformations is satisfied. 

This is a significant result, as it shows that the restricted volume-preserving
deformation retains the necessary algebraic structure for consistent
matrix regularization while maintaining Lorentz covariance.

In this way, by replacing the Poisson bracket with commutators and
performing matrix regularization on the gauge-fixed action, we obtain:

\begin{equation}
S=\int d^{3}\sigma\frac{1}{2}\left([\tau(X^{I},X^{J}),X^{K}]+\frac{\partial X^{I}}{\partial\sigma^{3}}[X^{J},X^{K}]+\Sigma(X^{I},X^{J};X^{K})\right)^{2},
\end{equation}
with the gauge-fixing condition:

\begin{equation}
C_{I}\frac{\partial X^{I}}{\partial\sigma^{3}}=\sigma^{3}.
\end{equation}
The Restricted Volume-Preserving Deformation (RVPD) in its matrix-regularized
form is expressed as:
\begin{equation}
\delta_{R}X^{I}=[\tau(Q_{1}^{(R)},Q_{2}^{(R)}),X^{I}].
\end{equation}
In the context of commutation relations, the RVPD exhibits the following
properties:
\begin{itemize}
\item \textbf{Linearity}:
\begin{equation}
\begin{aligned}[][\tau(Q_{1}^{(R)},Q_{2}^{(R)}),\lambda_{1}A_{1}+\lambda_{2}A_{2}+\dots] & =\lambda_{1}[\tau(Q_{1}^{(R)},Q_{2}^{(R)}),A_{1}]+\dots\\
 & +\lambda_{n}[\tau(Q_{1}^{(R)},Q_{2}^{(R)}),A_{n}]
\end{aligned}
\end{equation}
\item \textbf{Distributive Property}:
\begin{equation}
\begin{aligned}[][\tau(Q_{1}^{(R)},Q_{2}^{(R)}),A_{1}A_{2}\dots] & =[\tau(Q_{1}^{(R)},Q_{2}^{(R)}),A_{1}]A_{2}\dots\\
 & +A_{1}[\tau(Q_{1}^{(R)},Q_{2}^{(R)}),A_{2}]\dots+\dots
\end{aligned}
\end{equation}
\item \textbf{Antisymmetry}:
\begin{equation}
[\tau(Q_{1}^{(R)},Q_{2}^{(R)}),A]=-[\tau(Q_{2}^{(R)},Q_{1}^{(R)}),A]
\end{equation}
\item \textbf{Conservation Law}:
\begin{equation}
[\tau(Q_{1}^{(R)},Q_{2}^{(R)}),Q_{1,2}^{(R)}]=0
\end{equation}
\item \textbf{Composition Rule of Transformations}:
\begin{equation}
\begin{aligned}[][\tau(Q_{1}^{(R)} & ,Q_{2}^{(R)}),[\tau(H_{1}^{(R)},H_{2}^{(R)}),A]]\\
 & =[\tau([\tau(Q_{1}^{(R)},Q_{2}^{(R)}),H_{1}^{(R)}],H_{2}^{(R)}),A]+[\tau(H_{1},[\tau(Q_{1}^{(R)},Q_{2}^{(R)}),H_{2}]),A]\\
 & +[\tau(H_{1}^{(R)},H_{2}^{(R)}),[\tau(Q_{1}^{(R)},Q_{2}^{(R)}),A]].
\end{aligned}
\end{equation}
\end{itemize}
Since $\delta_{R}X^{I}$ still satisfies the composition rule of transformations
even when expressed using commutators, the matrix-regularized version
of RVPD is preserved as a symmetry of the action.

Furthermore, the resulting matrix model remains globally Lorentz invariant.
Thus, a Lorentz-invariant matrix model for membranes is successfully
derived from the matrix regularization of the membrane action while
maintaining the essential RVPD symmetry.

\section{The equations of motion}

The variation of the obtained matrix model action is given by:

\begin{equation}
\begin{aligned}\delta S & =\int d\sigma^{3}\mathrm{Tr}\epsilon_{ijk}\epsilon_{i'j'k'}[X^{I_{i}},X^{I_{j}},X^{I_{k}}][\delta\tau(X^{I_{i'}},X^{I_{j'}}),X^{I_{k'}}]\\
 & +\int d\sigma^{3}\mathrm{Tr}\epsilon_{ijk}\epsilon_{i'j'k'}[X^{I_{i}},X^{I_{j}},X^{I_{k}}][\tau(X^{I_{i'}},X^{I_{j'}}),\delta X^{I_{k'}}].
\end{aligned}
\end{equation}
After computing this variation, we derive the equation of motion:

\begin{equation}
\begin{aligned}\epsilon_{ijk}\epsilon_{i'j'k'}\left([\tau(X^{I_{k'}},X^{I_{j'}}),[X^{I_{k}},X^{I_{j}},X^{I_{i}}]]\right.\\
-2\frac{\partial}{\partial\sigma^{3}}\left(X^{I_{j'}}[X^{I_{k'}},[X^{I_{k}},X^{I_{i}},X^{I_{j}}]]\right)\bigr) & =0.
\end{aligned}
\end{equation}

This equation of motion encapsulates the dynamics of the matrix model
derived from the Lorentz-covariant formulation of the membrane action.
The appearance of the commutator and the specific form of the volume-preserving
deformation demonstrate how the restricted volume-preserving symmetry
(RVPD) influences the dynamics of the membrane's matrix regularization.

\subsection{Solutions}

The equation of motion admits solutions of the form:

\begin{equation}
\begin{aligned}\epsilon_{ijk}[X^{I_{i}},X^{I_{j}},X^{I_{k}}]= & \frac{1}{3!}\epsilon_{ijk}[\tau(X^{I_{i}},X^{I_{j}}),X^{I_{k}}]+\frac{1}{3!}\epsilon_{ijk}\frac{\partial}{\partial\sigma^{3}}\left(X^{I_{i}}[X^{I_{j}},X^{I_{k}}]\right)\\
 & =g^{I_{1}I_{2}I_{3}}(\sigma^{3}),
\end{aligned}
\end{equation}
where $g^{I_{1}I_{2}I_{3}}(\sigma^{3})$ is a function of $\sigma^{3}$.
This form satisfies the equations of motion under specific conditions
on the matrices $X^{I}$.

\subsection{Particle-like Solutions}

A specific example of a solution is the particle-like configuration:

\begin{equation}
X^{0}=\sigma^{3},
\end{equation}

\begin{equation}
X^{1,\dots,10}=f^{1,\dots,10}(\sigma^{3}),
\end{equation}
where $f^{1,\dots,10}(\sigma^{3})$ are functions that satisfy the
gauge-fixing condition.

This solution represents a particle-like state in the matrix model,
where the spatial components of the membrane are governed by the functions
$f^{1,\dots,10}(\sigma^{3})$, and the temporal component is linearly
dependent on $\sigma^{3}$. The particle interpretation arises because
all spatial coordinates are reduced to functions of a single parameter,
$\sigma^{3}$, resembling a worldline rather than an extended membrane.

\subsection{Non-Commutative Membrane Solutions}

Another class of solutions is the non-commutative membrane configuration:
\begin{equation}
X^{0}=\sigma^{3},
\end{equation}

\begin{equation}
X^{1}=x^{1},
\end{equation}

\begin{equation}
X^{2}=x^{2},
\end{equation}

\begin{equation}
X^{3,\dots,10}=0,
\end{equation}
with the non-commutative relation:

\begin{equation}
[x^{1},x^{2}]=i\theta.
\end{equation}

This configuration describes a two-dimensional membrane (M2-brane)
that is extended in a non-commutative manner. The coordinates $x^{1}$
and $x^{2}$ obey the Heisenberg-like commutation relation, indicating
a quantum geometry in the membrane's spatial extension.

In this scenario, $\sigma^{3}$ plays the role of time, with the membrane
evolving over time while maintaining its non-commutative structure
in the two spatial dimensions. This non-commutative solution provides
a concrete example of how the matrix model can describe extended objects
with inherent quantum geometric properties.

\subsection{Multiple Non-Commutative Membranes}

To describe multiple non-commutative membranes, one can use block-diagonal
matrices. For example, in the case of two membranes, the matrices
can be chosen as:
\begin{equation}
X^{1}=\left(\begin{array}{cc}
x^{1} & 0\\
0 & x^{1}
\end{array}\right),
\end{equation}

\begin{equation}
X^{2}=\left(\begin{array}{cc}
x^{2} & 0\\
0 & x^{2}
\end{array}\right).
\end{equation}
This configuration effectively represents two non-commutative membranes
that do not interact, as the matrices are block-diagonal with identical
elements.

To introduce interactions between the membranes, one can add off-diagonal
elements to the matrices\footnote{For the method of introducing interactions into diagonal terms, see
for example \citep{Ishibashi_1997}.}. For example:
\begin{equation}
X^{1}=\left(\begin{array}{cc}
x^{1} & \phi\\
\phi & x^{1}
\end{array}\right).
\end{equation}
Here, $\phi$ represents the interaction between the two membranes.
This approach is analogous to how interactions are modeled in other
matrix models, where off-diagonal elements mediate the dynamics between
different branes or extended objects.

By tuning the off-diagonal terms, one can control the strength and
nature of the interaction, allowing for the modeling of phenomena
such as brane collisions, bound states, or dynamic exchanges of energy
and momentum between membranes.

\subsection{4, 6, 8, 10-Dimensional Non-Commutative Membranes}

A 4-dimensional non-commutative membrane can be constructed using
the following configuration:

\begin{equation}
X^{0}=\sigma^{3},
\end{equation}

\begin{equation}
X^{1}=x^{1},
\end{equation}

\begin{equation}
X^{2}=x^{2},
\end{equation}

\begin{equation}
X^{3}=x^{1},
\end{equation}

\begin{equation}
X^{4}=x^{2},
\end{equation}

\begin{equation}
X^{5,\dots,10}=0,
\end{equation}
with the non-commutative relation:

\begin{equation}
[x^{1},x^{2}]=i\theta.
\end{equation}
This setup effectively ``duplicates'' the non-commutative plane across
additional dimensions, resulting in a 4-dimensional non-commutative
membrane. The method can be extended similarly to create 6, 8, and
10-dimensional non-commutative membranes by introducing more such
pairs of spatial coordinates while maintaining the same non-commutative
relation between each pair.

For a 6-dimensional membrane, we might set:

\begin{equation}
X^{5}=x^{1},\ X^{6}=x^{2}.
\end{equation}
And similarly, for 8 and 10 dimensions, more coordinates can be assigned
in the same manner. The extension to higher dimensions maintains the
structure of the non-commutative geometry by ensuring that the same
commutation relations hold between the appropriate coordinate pairs.
This method provides a systematic way to construct higher-dimensional
non-commutative branes within the matrix model framework while preserving
the symmetry and algebraic consistency of the model.

\section{Conclusion}

In this study, a Lorentz covariant matrix model was obtained by partially
restricting the volume-preserving deformation (VPD) through gauge
fixing, resulting in a Restricted Volume-Preserving Deformation (RVPD)
within the bosonic part of the M2-brane action in 11-dimensional spacetime.

We demonstrated that the solutions to this matrix model include particle-like
solutions, two-dimensional non-commutative membranes, as well as higher-dimensional
non-commutative membranes with 4, 6, 8, and 10 dimensions. These results
suggest that the proposed matrix model is capable of describing a
wide variety of extended objects within a consistent Lorentz-invariant
framework.

\section{Discussion}

To assess whether these solutions are stable, it is essential to incorporate
supersymmetry into the model. Supersymmetry could provide the necessary
framework to analyze stability and identify potential BPS states within
the matrix model.

Additionally, demonstrating the correspondence between this Lorentz
covariant matrix model and conventional discussions of M2-branes,
such as those in the BFSS model or the BLG model, is crucial. Establishing
such connections would not only validate the proposed model but also
facilitate comparisons with established non-perturbative formulations
of M-theory.

Since M5-branes can also be described using higher-order Nambu brackets,
extending the current analysis to M5-branes is a promising direction.
Such an extension could potentially reveal new insights into the non-perturbative
structure of M-theory and offer a unified approach to describing multiple
types of branes.

This study establishes a robust foundation for future investigations,
including the exploration of supersymmetry, model correspondences,
and higher-dimensional brane theories within the Lorentz covariant
matrix model framework. In this regard, the author is currently developing
a supersymmetric extension of the model, to be reported in a forthcoming
publication\footnote{
A supersymmetric extension of the present bosonic model, incorporating $\kappa$-symmetry and a systematic classification of BPS states, has already been accepted for publication in \textit{JHEP} (see \citep{katagiriSusy}). 
This demonstrates that the RVPD-based matrix regularization framework consistently accommodates both bosonic and supersymmetric M2-branes.}.

\section*{Acknowledgments}

First and foremost, I am also deeply grateful to Professor Emeritus
Akio Sugamoto of Ochanomizu University for his constant warm encouragement
and for offering numerous important insights during this period.

I sincerely appreciate Associate Professor Shiro Komata of the Open
University of Japan for reviewing my manuscript and providing valuable
comments on its structure and presentation. I would also like to thank
Mr. Noriaki Aibara for checking the references.

Finally, with deep gratitude and remembrance, I dedicate this paper
to the memory of my late friend, Mr. Wakata, with whom I was supposed
to graduate when I withdrew from the University of Tsukuba's graduate
program twenty years ago.

\appendix
\section*{Appendices}
\addcontentsline{toc}{section}{Appendices}

\section{Algebraic Aspects of RVPD}

In this appendix, we discuss the algebraic aspects of the Restricted
Volume-Preserving Deformation (RVPD). 

The RVPD forms an algebra defined by the following relations:

\begin{equation}
\tau(Q_{1},Q_{2})\equiv\frac{\partial Q_{1}}{\partial\sigma^{3}}Q_{2}-\frac{\partial Q_{2}}{\partial\sigma^{3}}Q_{1},
\end{equation}

\begin{equation}
[Q_{1},Q_{2}]=0,
\end{equation}

\begin{equation}
\frac{\partial}{\partial\sigma^{3}}\tau(Q_{1},Q_{2})=0.
\end{equation}
For any arbitrary element $A$, the following condition holds:

\begin{equation}
[\frac{\partial}{\partial\sigma^{3}}Q_{1,2},A]=0.
\end{equation}
These algebraic conditions define the structure of the RVPD, ensuring
that the restricted transformations preserve the necessary composition
rules and maintain the consistency of the matrix model under the Lorentz
covariant framework.

\subsection*{A.1 Commutation Relations of RVPD Generators}

The algebra of the RVPD can be analyzed using the commutation relations
of the generators:

\begin{equation}
[\tau(Q_{1},Q_{2}),\tau(H_{1},H_{2})]=\tau([\tau(Q_{1},Q_{2}),H_{1}],H_{2})+\tau(H_{1},[\tau(Q_{1},Q_{2}),H_{2}]).
\end{equation}

Using this relation, we can define the action of the RVPD on an arbitrary
element $A$ as:
\begin{equation}
\delta_{R(Q_{1},Q_{2})}A\equiv[\tau(Q_{1},Q_{2}),A].
\end{equation}
The composition of two RVPD transformations is given by:

\begin{equation}
\delta_{R(Q_{1},Q_{2})}\delta_{R(H_{1},H_{2})}A=[\tau(Q_{1},Q_{2}),[\tau(H_{1},H_{2}),A]].
\end{equation}
By applying the Jacobi identity and the commutation relations, this
expands to:

\begin{equation}
=[[\tau(Q_{1},Q_{2}),\tau(H_{1},H_{2})],A]+[\tau(H_{1},H_{2}),[\tau(Q_{1},Q_{2}),A]].
\end{equation}
This can be further rewritten as:

\begin{equation}
=\delta_{R(\delta_{R(Q_{1},Q_{2})}H_{1},H_{2})}A+\delta_{R(H_{1},\delta_{R(Q_{1},Q_{2})}H_{2})}A+\delta_{R(H_{1},H_{2})}\delta_{R(Q_{1},Q_{2})}A.
\end{equation}

\subsection*{A.2 Decomposition of $Q_{1}$ and $Q_{2}$}

Given the constraints on $Q_{1},Q_{2}$, they can be decomposed as:
\begin{equation}
Q_{1,2}=\phi_{1,2}^{(Q)}(\sigma^{1},\sigma^{2})+\chi_{1.2}^{(Q)}(\sigma^{3}).
\end{equation}
The commutativity condition:

\begin{equation}
[Q_{1},Q_{2}]=0
\end{equation}
implies that:

\begin{equation}
[\phi_{1}^{(Q)},\phi_{2}^{(Q)}]=0.
\end{equation}

\subsection*{A.3 Expression of $\tau(Q_{1},Q_{2})$}

The generator $\tau(Q_{1},Q_{2})$ can be expressed as:
\begin{equation}
\tau(Q_{1},Q_{2})=\frac{\partial\chi_{1}^{(Q)}(\sigma^{3})}{\partial\sigma^{3}}\phi_{2}^{(Q)}(\sigma^{1},\sigma^{2})-\frac{\partial\chi_{2}^{(Q)}(\sigma^{3})}{\partial\sigma^{3}}\phi_{1}^{(Q)}(\sigma^{1},\sigma^{2}).
\end{equation}

For consistency with:

\begin{equation}
\frac{\partial\tau}{\partial\sigma^{3}}(Q_{1},Q_{2})=0,
\end{equation}
we derive the conditions:
\begin{itemize}
\item When $\phi_{1}=\phi_{2}$,
\end{itemize}
\begin{equation}
\left(\frac{\partial}{\partial\sigma^{3}}\right)^{2}\chi_{1}^{(Q)}(\sigma^{3})=\left(\frac{\partial}{\partial\sigma^{3}}\right)^{2}\chi_{2}^{(Q)}(\sigma^{3})
\end{equation}

\begin{itemize}
\item When $\phi_{1}\neq\phi_{2}$,
\begin{equation}
\left(\frac{\partial}{\partial\sigma^{3}}\right)^{2}\chi_{1,2}^{(Q)}(\sigma^{3})=0.
\end{equation}
\end{itemize}

\subsection*{A.4 Commutation of Generators}

For elements $H_{1},H_{2}$ with similar properties, we also have:
\begin{equation}
H_{1,2}=\phi_{1,2}^{(H)}(\sigma^{1},\sigma^{2})+\chi_{1,2}^{(H)}(\sigma^{3}).
\end{equation}
Calculating the commutator of the RVPD generators:

\begin{equation}
[\tau(Q_{1},Q_{2}),\tau(H_{1},H_{2})]
\end{equation}
leads to:

\begin{equation}
\begin{aligned}= & \frac{\partial\chi_{1}^{(Q)}(\sigma^{3})}{\partial\sigma^{3}}\frac{\partial\chi_{1}^{(H)}(\sigma^{3})}{\partial\sigma^{3}}[\phi_{2}^{(Q)}(\sigma^{1},\sigma^{2}),\phi_{2}^{(H)}(\sigma^{1},\sigma^{2})]\\
 & +\frac{\partial\chi_{2}^{(Q)}(\sigma^{3})}{\partial\sigma^{3}}\frac{\partial\chi_{2}^{(H)}(\sigma^{3})}{\partial\sigma^{3}}[\phi_{1}^{(Q)}(\sigma^{1},\sigma^{2}),\phi_{1}^{(H)}(\sigma^{1},\sigma^{2})]\\
 & -\frac{\partial\chi_{1}^{(Q)}(\sigma^{3})}{\partial\sigma^{3}}\frac{\partial\chi_{2}^{(H)}(\sigma^{3})}{\partial\sigma^{3}}[\phi_{2}^{(Q)}(\sigma^{1},\sigma^{2}),\phi_{1}^{(H)}(\sigma^{1},\sigma^{2})]\\
 & -\frac{\partial\chi_{2}^{(Q)}(\sigma^{3})}{\partial\sigma^{3}}\frac{\partial\chi_{1}^{(H)}(\sigma^{3})}{\partial\sigma^{3}}[\phi_{1}^{(Q)}(\sigma^{1},\sigma^{2}),\phi_{2}^{(H)}(\sigma^{1},\sigma^{2})].
\end{aligned}
\end{equation}

\subsection*{A.5 Classification of $\phi(\sigma^{1},\sigma^{2})$}

The functions $\phi(\sigma^{1},\sigma^{2})$ can be classified into
sets $\Sigma_{1},\Sigma_{2},\dots$as mutually commutative elements
such as $\left(\sigma^{1}\right)^{n}\left(\sigma^{2}\right)^{m}$.

If $Q_{1,2}^{(a)}\in\Sigma_{a}$, then:
\begin{equation}
[Q_{1}^{(a)},Q_{2}^{(a)}]=0,
\end{equation}

\begin{equation}
[Q_{i}^{(a)},Q_{j}^{(b)}]=f_{ijk}^{(abc)}Q_{k}^{(c)}.
\end{equation}
For example:
\begin{equation}
\Sigma_{1}=\{\left(\sigma^{1}\right)^{n}\}_{n},
\end{equation}

\begin{equation}
\Sigma_{2}=\{\left(\sigma^{1}\right)^{n}\left(\sigma^{2}\right)^{n}\}_{n}.
\end{equation}

\begin{equation}
[\left(\sigma^{1}\right)^{n},\sigma^{1}\sigma^{2}]=n\left(\sigma^{1}\right)^{n}.
\end{equation}

\subsection*{A.6 Interpretation of the Algebra}

From the above, this algebra has the properties of an \textit{infinite-dimensional
Lie algebra} and can be decomposed as a \textit{direct sum of commutative
subalgebras}. At least for the non-trivial part, it has a \textit{Witt
algebra} structure. The potential relation to the $w(\infty)$ algebra
is also an interesting direction for further investigation, as it
may reveal deeper connections with symmetries of higher-dimensional
branes or extended objects in M-theory.

\section{Detailed Derivation of the Gauge Parameter Constraints (RVPD) from
the Gauge Restriction Condition}

This appendix provides the technical details of how the gauge parameter
constraints (RVPD) are derived from the gauge restriction condition.

\subsection*{B.1 Preliminaries}

We consider a membrane in 11-dimensional spacetime (two spatial dimensions
and one temporal dimension).

After an appropriate gauge fixing of the Nambu--Goto action, the
bosonic part of the action can be written as

\begin{equation}
S=\int d^{3}\sigma\frac{1}{2}\{X^{I},X^{J},X^{K}\}^{2}.
\end{equation}
Here, $X^{I}=X^{I}(\sigma^{1},\sigma^{2},\sigma^{3})$, where $\sigma^{1},\sigma^{2},\sigma^{3}$
are coordinates on the worldvolume of the membrane.

From here on, the index $i=1,2,3$ refers to the worldvolume coordinates
$\sigma^{i}$, and the index $I=0,\dots,10$ refers to the spacetime
coordinates $X^{I}$. 

The expression $\{X^{I},X^{J},X^{K}\}$ denotes the Nambu bracket,
defined by
\begin{equation}
\{X^{I},X^{J},X^{K}\}\equiv\epsilon^{ijk}\frac{\partial X^{I}}{\partial\sigma^{i}}\frac{\partial X^{J}}{\partial\sigma^{j}}\frac{\partial X^{K}}{\partial\sigma^{k}}.
\end{equation}
The membrane action is invariant under volume-preserving deformations
(VPD) of the form

\begin{equation}
\delta_{VPD}X^{I}\equiv\{Q_{1},Q_{2},X^{I}\}
\end{equation}
\foreignlanguage{english}{where $Q_{1}$ and $Q_{2}$ are arbitrary
gauge parameters depending on $(\sigma^{1},\sigma^{2},\sigma^{3})$. }

We assume that the gauge parameters are sufficiently smooth and free
of singularities with respect $\sigma^{3}$. On the other hand, we
allow for some degree of local discontinuities or zeros with respect
to $\sigma^{1}$ and $\sigma^{2}$\footnote{This is because $\sigma^{3}$ plays the role of a time-like or evolution
parameter, as discussed in the equations of motion in the main body
of the paper. }. We also assume that appropriate boundary conditions are imposed
on the gauge parameters.

In this paper, we decompose the Nambu bracket into the Poisson bracket
with respect to $\sigma^{1}$ and $\sigma^{2}$ as follows:

\begin{equation}
\{X^{I},X^{J},X^{K}\}=\{\tau(X^{I},X^{J}),X^{K}\}+\frac{\partial X^{K}}{\partial\sigma^{3}}\{X^{I},X^{J}\}+\Sigma(X^{I},X^{J};X^{K})
\end{equation}
where $\{X^{I},X^{J}\}$ denotes the Poisson bracket with respect
to $\sigma^{1}$ and $\sigma^{2}$, defined by

\begin{equation}
\{X^{I},X^{J}\}\equiv\epsilon^{ab}\frac{\partial X^{I}}{\partial\sigma^{a}}\frac{\partial X^{J}}{\partial\sigma^{b}},
\end{equation}
with $a=1,2$. The functions $\Sigma(A,B;C)$ and $\tau(A,B)$ are
defined as 
\begin{equation}
\Sigma(A,B;C)\equiv A\{\frac{\partial B}{\partial\sigma^{3}},C\}-B\{\frac{\partial A}{\partial\sigma^{3}},C\},
\end{equation}

\begin{equation}
\tau(A,B)\equiv\frac{\partial A}{\partial\sigma^{3}}B-\frac{\partial B}{\partial\sigma^{3}}A.
\end{equation}
Both of these expressions are antisymmetric with respect to $A$ and
$B$. This decomposition is simply an equivalent rewriting of the
Nambu bracket.

Using this decomposition, the volume-preserving deformation (VPD)
can be expressed as

\begin{equation}
\delta_{\mathrm{VPD}}X^{I}=\{\tau(Q_{1},Q_{2}),X^{I}\}+\frac{\partial X^{I}}{\partial\sigma^{3}}\{Q_{1},Q_{2}\}+\Sigma(Q_{1},Q_{2};X^{I}).
\end{equation}

\subsection*{B.2 Main Claim}

We impose the ``gauge restriction condition''\footnote{In this work, we use the term \textquotedbl gauge restriction condition\textquotedbl{}
instead of \textquotedbl gauge fixing condition.\textquotedbl{} This
is because it does not refer to the elimination of redundant degrees
of freedom in the usual sense, but rather denotes an algebraic constraint
required for the consistency of the theoretical construction.}\footnote{More generally, the gauge restriction condition can be written as
\foreignlanguage{japanese}{$C_{I}\frac{\partial X^{I}}{\partial\sigma^{3}}=h(\sigma^{3})$
, where $h(\sigma^{3})$ is a monotonic function. In this paper, we
simply take $h(\sigma^{3})=\sigma^{3}$for convenience.} }
\begin{equation}
C_{I}\frac{\partial X^{I}}{\partial\sigma^{3}}=\sigma^{3},
\end{equation}
where $C_{I}$ is a fixed Lorentz vector in spacetime.

By integrating the gauge restriction condition, we obtain
\begin{equation}
C_{I}X^{I}=\frac{1}{2}(\sigma^{3})^{2}+f(\sigma^{1},\sigma^{2})
\end{equation}
where $f(\sigma^{1},\ \sigma^{2})$ appears as an integration constant
along the $\sigma^{3}$ direction. As long as it is sufficiently smooth
and satisfies appropriate boundary conditions, it can take arbitrary
values.

We require that the right-hand side of the equation, including $f(\sigma^{1},\sigma^{2})$,
remain invariant under volume-preserving deformations (VPD)\footnote{The fact that $f(\sigma^{1},\sigma^{2})$ is arbitrary---provided
it is sufficiently smooth and satisfies appropriate boundary conditions---plays
a key role in deriving strong constraints such as $\{Q_{1},Q_{2}\}=0$
later in the analysis. In the following sections, we carefully track
where and how this arbitrariness is used throughout the calculations.}\footnote{In this work, we impose strong constraints on the gauge parameters
$Q_{1}$ and $Q_{2}$ in order to preserve \textit{all} possible choices
of $f(\sigma_{1},\sigma_{2})$. This is not merely a technical requirement
to ensure consistency of the gauge condition, but rather a necessary
condition to define the algebraic structure and composition law of
the restricted volume-preserving deformation (RVPD) in a precise and
consistent manner. In particular, as shown in the main text, under
the conditions such as $\{Q_{1},Q_{2}\}=0$ and $\partial_{\sigma^{3}}\tau(Q_{1},Q_{2})=0$,
the deformation closes algebraically, ensuring that the structure
remains intact even after quantization or matrix regularization. Thus,
the very policy of preserving all choices of $f$ becomes the key
to achieving consistency between the algebraic structure and physical
content of the theory. In this way, the requirement that the gauge
condition be preserved for every $f$ should not be viewed as a demand
to leave $f$ unchanged, but rather as a natural condition to obtain
a closed deformation algebra corresponding to RVPD.}\footnote{It is important to emphasize that, in the expression\foreignlanguage{japanese}{
$C_{I}X^{I}=\frac{1}{2}(\sigma^{3})^{2}+f(\sigma^{1},\sigma^{2})$
used in this work, we are not choosing a particular function $f_{0}$
for the purpose of gauge fixing. Instead, $f(\sigma^{1},\sigma^{2})$
is treated as an arbitrary allowed function, and only those gauge
transformations that preserve \textit{all} such possible functions
are considered. This is the reason why strong constraints such as
$\{Q_{1},Q_{2}\}=0$ arise. As noted earlier, these constraints play
a crucial role in maintaining stability under quantization (i.e.,
matrix regularization) and preserving the underlying algebraic structure.}}\footnote{Note that any specific solution $X^{I}(\sigma^{1},\sigma^{2},\sigma^{3})$
satisfies the gauge condition $C_{I}X^{I}=\frac{1}{2}(\sigma^{3})^{2}+f(\sigma^{1},\sigma^{2})$
for \textit{some} particular choice of $f$, but not for all $f$
simultaneously. In this paper, we only allow deformations (i.e., RVPD)
that preserve all possible $f$, so that the physical structure remains
consistent regardless of which gauge-fixing surface (i.e., which $f$)
a solution belongs to. The function $f$ associated with a given solution
is determined by the initial or boundary conditions of $X^{I}$, and
the theoretical framework is constructed under the assumption that
$f$ is arbitrary. }. On the other hand, we require that the gauge parameters $Q_{1}$
and $Q_{2}$ do not depend on the integration constant $f$\footnote{This is because, if $Q$ depends on $f$, the argument for preserving
all possible $f$ simultaneously would become self-referential and
logically circular.}.

Under this requirement, the volume-preserving deformation is subject
to the following constraints:

\begin{equation}
\frac{\partial\tau(Q_{1},Q_{2})}{\partial\sigma^{3}}=0,
\end{equation}

\begin{equation}
\{Q_{1},Q_{2}\}=0,
\end{equation}

\begin{equation}
\frac{\partial}{\partial\sigma^{3}}\frac{\partial Q_{1,2}}{\partial\sigma^{a}}=0.
\end{equation}
These constraints on the gauge parameters are both necessary and sufficient
conditions for preserving the gauge restriction condition.

We refer to the volume-preserving deformations that satisfy these
constraints as \textit{Restricted Volume-Preserving Deformations (RVPD)}.

As a result, under RVPD, the terms $\{Q_{1},Q_{2}\}$ and $\Sigma(Q_{1},Q_{2};X^{I})$
in the decomposition of the Nambu bracket vanish, and the deformation
takes the simplified form

\begin{equation}
\delta_{R}X^{I}=\{\tau(Q_{1},Q_{2}),X^{I}\}.
\end{equation}

The physical meaning of each of the parameter constraints is as follows:
\begin{itemize}
\item $\{Q_{1},Q_{2}\}=0$: This indicates that the Poisson bracket on the
$(\sigma^{1},\sigma^{2})$ plane vanishes, implying that $Q_{1}$
and $Q_{2}$ are locally dependent on each other in the two-dimensional
base space.
\item $\partial_{\sigma^{3}}\tau(Q_{1},Q_{2})=0$: This implies that $\tau(Q_{1},Q_{2})$
is independent of $\sigma^{3}$, and can therefore be regarded as
a constant along the $\sigma^{3}$ direction.
\item $\partial_{\sigma^{3}}\partial_{a}Q_{1,2}=0$: This condition ensures
that the $(\sigma^{1},\sigma^{2})$-dependence of $Q_{1}$ and $Q_{2}$
remains unchanged under differentiation with respect to $\sigma^{3}$.
\end{itemize}
These constraints can be interpreted as strong restrictions on the
$\sigma^{3}$-dependence of the gauge parameters, imposed to prevent
large variations in the $\sigma^{3}$ direction from violating the
gauge restriction condition. In particular, the condition $\{Q_{1},Q_{2}\}=0$
is the most essential one for ensuring that the arbitrary function
$f(\sigma^{1},\sigma^{2})$ remains unchanged.

\subsection*{B.2.1 Comparison with Conventional Gauge Fixing Conditions}

In a conventional gauge fixing procedure, one selects a particular
choice of $f$, thereby fixing the gauge freedom by restricting the
system to a single geometric surface (gauge-fixing surface) corresponding
to that specific value.

In contrast, the \textit{gauge restriction condition} treated in this
work considers the entire family of such surfaces corresponding to
all possible choices of $f$, and requires that none of them be altered
under gauge transformations.

As a result, constraints such as $\{Q_{1},Q_{2}\}=0$ naturally emerge,
and the resulting structure is qualitatively different from that of
ordinary gauge fixing.

This approach is essential in our framework in order to consistently
treat the Nambu bracket, volume-preserving deformations, and quantization
(matrix regularization)\footnote{In this work, we adopt this approach in order to treat the Nambu bracket,
volume-preserving deformations, and quantization (matrix regularization)
in a consistent manner. However, this does not preclude the possibility
of other approaches. The existence of alternative methods for quantizing
the Nambu bracket or performing matrix regularization lies beyond
the scope of this study, and the discussion remains open.}.

\subsection*{B.3 Proof of Sufficiency}

We first show that, as a sufficient condition, if the constraints
on the gauge parameters are satisfied, then the gauge restriction
condition is preserved.

That is, starting from the gauge restriction condition

\begin{equation}
C_{I}\frac{\partial X^{I}}{\partial\sigma^{3}}=\sigma^{3},
\end{equation}
we apply an RVPD transformation and examine whether the condition
remains preserved:

\begin{equation}
C_{I}\frac{\partial\delta_{R}X^{I}}{\partial\sigma^{3}}=0.
\end{equation}
Since RVPD satisfies\foreignlanguage{english}{ $\frac{\partial\tau(Q_{1},Q_{2})}{\partial\sigma^{3}}=0$,
we have}
\begin{equation}
C_{I}\{\tau(Q_{1},Q_{2}),\frac{\partial X^{I}}{\partial\sigma^{3}}\}=0.
\end{equation}
This implies
\begin{equation}
\{\tau(Q_{1},Q_{2}),C_{I}\frac{\partial X^{I}}{\partial\sigma^{3}}\}=0,
\end{equation}
and therefore
\begin{equation}
\{\tau(Q_{1},Q_{2}),\sigma_{3}\}=0,
\end{equation}
which confirms that the gauge restriction condition is indeed preserved.

\subsection*{B.4 Proof of Necessity}

We begin by applying a volume-preserving deformation (VPD) to the
gauge restriction condition, which leads to the requirement
\[
C_{I}\frac{\partial\delta_{VPD}X^{I}}{\partial\sigma^{3}}=0.
\]

Here, we have\foreignlanguage{english}{
\begin{equation}
\delta_{\mathrm{VPD}}X^{I}=\{\tau(Q_{1},Q_{2}),X^{I}\}+\frac{\partial X^{I}}{\partial\sigma^{3}}\{Q_{1},Q_{2}\}+\Sigma(Q_{1},Q_{2};X^{I}).
\end{equation}
}

Substituting this into the previous expression, we obtain:

\[
\begin{aligned}\{\frac{\partial\tau(Q_{1},Q_{2})}{\partial\sigma^{3}},X^{I}\}+C_{I}\{\tau(Q_{1},Q_{2}),C_{I}\frac{\partial X^{I}}{\partial\sigma^{3}}\}\\
+\frac{\partial}{\partial\sigma^{3}}\left(C_{I}\frac{\partial X^{I}}{\partial\sigma^{3}}\right)\{Q_{1},Q_{2}\}+C_{I}\frac{\partial X^{I}}{\partial\sigma^{3}}\frac{\partial}{\partial\sigma^{3}}\{Q_{1},Q_{2}\}\\
+C_{I}\Sigma(\frac{\partial}{\partial\sigma^{3}}Q_{1},Q_{2};X^{I})+C_{I}\Sigma(Q_{1},\frac{\partial}{\partial\sigma^{3}}Q_{2};X^{I})\\
+\Sigma(Q_{1},Q_{2};C_{I}\frac{\partial}{\partial\sigma^{3}}X^{I}) & =0.
\end{aligned}
\]

Now, using the gauge restriction condition $C_{I}\frac{\partial}{\partial\sigma^{3}}X^{I}=\sigma^{3}$,
this expression simplifies to:
\begin{equation}
\begin{aligned}\{\frac{\partial\tau(Q_{1},Q_{2})}{\partial\sigma^{3}},C_{I}X^{I}\}+\{Q_{1},Q_{2}\}+\sigma^{3}\frac{\partial}{\partial\sigma^{3}}\{Q_{1},Q_{2}\}\\
+\Sigma(\frac{\partial}{\partial\sigma^{3}}Q_{1},Q_{2};C_{I}X^{I})+\Sigma(Q_{1},\frac{\partial}{\partial\sigma^{3}}Q_{2};C_{I}X^{I}) & =0.
\end{aligned}
\end{equation}

As shown previously in Section B.2, integrating the gauge restriction
condition $C_{I}\frac{\partial X^{I}}{\partial\sigma^{3}}=\sigma^{3}$
along the $\sigma^{3}$ direction gives
\begin{equation}
C_{I}X^{I}=\frac{1}{2}\left(\sigma^{3}\right)^{2}+f(\sigma^{1},\sigma^{2})
\end{equation}
where $f(\sigma^{1},\sigma^{2})$ is an arbitrary integration constant.

We now substitute this expression into the bracket relations to proceed
with the proof.

Substituting this result, we note that since $C_{I}X^{I}$ appears
inside the Poisson bracket, the quadratic term $\frac{1}{2}(\sigma^{3})^{2}$
drops out. Thus, the equation becomes:

\begin{equation}
\begin{aligned}\{\frac{\partial\tau(Q_{1},Q_{2})}{\partial\sigma^{3}},f(\sigma^{1},\sigma^{2})\}+\{Q_{1},Q_{2}\}+\sigma^{3}\frac{\partial}{\partial\sigma^{3}}\{Q_{1},Q_{2}\}\\
+\Sigma(\frac{\partial}{\partial\sigma^{3}}Q_{1},Q_{2};f(\sigma^{1},\sigma^{2}))+\Sigma(Q_{1},\frac{\partial}{\partial\sigma^{3}}Q_{2};f(\sigma^{1},\sigma^{2})) & =0.
\end{aligned}
\end{equation}
Here, since $f(\sigma^{1},\sigma^{2})$ is arbitrary, each term involving
$f$ must vanish independently. Therefore, the following two conditions
must hold:
\begin{equation}
\{Q_{1},Q_{2}\}+\sigma^{3}\frac{\partial}{\partial\sigma^{3}}\{Q_{1},Q_{2}\}=0,
\end{equation}

\begin{equation}
\Sigma(\frac{\partial}{\partial\sigma^{3}}Q_{1},Q_{2};f(\sigma^{1},\sigma^{2}))+\Sigma(Q_{1},\frac{\partial}{\partial\sigma^{3}}Q_{2};f(\sigma^{1},\sigma^{2}))=0.
\end{equation}
The first equation is a differential equation with respect to $\sigma^{3}$.
Solving it yields: 
\begin{equation}
\{Q_{1},Q_{2}\}=\frac{C(\sigma^{1},\sigma^{2})}{\sigma^{3}},
\end{equation}
where $C(\sigma^{1},\sigma^{2})$ is an integration constant in the
$\sigma^{3}$ direction.

However, since we have assumed that $Q_{1}$ and $Q_{2}$ are sufficiently
smooth functions of $\sigma^{3}$, the appearance of $\sigma^{3}$
in the denominator would introduce a singularity at $\sigma^{3}=0$,
which is not allowed under our regularity assumption in the $\sigma^{3}$direction.
Even though local irregularities in $\sigma^{1}$, $\sigma^{2}$ are
permitted, such a singularity in $\sigma^{3}$ is incompatible with
the smoothness conditions imposed earlier. Therefore, we must have
$C(\sigma^{1},\sigma^{2})=0$, and we conclude\footnote{%
This regularity assumption arises naturally from physical considerations
such as boundary conditions or the requirement of a finite membrane
configuration.%
}\footnote{%
Note that this argument does not rely on the special case $f=0$.
Rather, it is precisely because $f$ is arbitrary that the terms independent
of $f$ must vanish.%
}:

\begin{equation}
\{Q_{1},Q_{2}\}=0.
\end{equation}
From this, the remaining equation becomes:

\begin{equation}
\begin{aligned}\{\frac{\partial\tau(Q_{1},Q_{2})}{\partial\sigma^{3}},f(\sigma^{1},\sigma^{2})\} & +\Sigma(\frac{\partial}{\partial\sigma^{3}}Q_{1},Q_{2};f(\sigma^{1},\sigma^{2}))\\
 & +\Sigma(Q_{1},\frac{\partial}{\partial\sigma^{3}}Q_{2};f(\sigma^{1},\sigma^{2}))=0.
\end{aligned}
\end{equation}
It should be noted that this equation cannot, in general, be written
in the form\foreignlanguage{english}{
\begin{equation}
\{A,f(\sigma_{1},\sigma_{2})\}=0
\end{equation}
for some function $A$.}

Therefore, we rewrite the equation by expanding the Poisson bracket,
yielding the following form:

\begin{equation}
K_{(\tau)}^{b}\partial_{b}f+K_{\Sigma1}^{b}\partial_{b}f+K_{\Sigma2}^{b}\partial_{b}f=0
\end{equation}
where
\begin{equation}
K_{(\tau)}^{b}\equiv\epsilon^{ab}\partial_{a}\partial_{\sigma^{3}}\tau(Q_{1},Q_{2}),
\end{equation}

\begin{equation}
K_{\Sigma1}^{b}\equiv\partial_{\sigma^{3}}Q_{1}\epsilon^{ab}\partial_{a}\partial_{\sigma^{3}}Q_{2}-Q_{2}\epsilon^{ab}\partial_{a}\partial_{\sigma^{3}}^{2}Q_{1},
\end{equation}

\begin{equation}
K_{\Sigma2}^{b}\equiv Q_{1}\epsilon^{ab}\partial_{a}\partial_{\sigma^{3}}^{2}Q_{2}-\partial_{\sigma^{3}}Q_{2}\epsilon^{ab}\partial_{a}\partial_{\sigma^{3}}Q_{1}.
\end{equation}
It should be emphasized that each $K^{b}$ is a coefficient of $\partial_{b}f$,
and---as is evident from the expressions above---is not itself a
differential operator.

The sum \foreignlanguage{japanese}{$K_{(\tau)}^{b}+K_{\Sigma1}^{b}+K_{\Sigma2}^{b}$
can be combined into the following equivalent expression:}

\begin{equation}
\begin{aligned}K_{(\tau)}^{b}+ & K_{\Sigma1}^{b}+K_{\Sigma2}^{b}=\\
 & \epsilon^{ab}\partial_{a}\partial_{\sigma^{3}}\tau(Q_{1},Q_{2})+\partial_{\sigma^{3}}\left(Q_{1}\epsilon^{ab}\partial_{\sigma^{3}}\partial_{a}Q_{2}-Q_{2}\epsilon^{ab}\partial_{\sigma^{3}}\partial_{a}Q_{1}\right)=0.
\end{aligned}
\end{equation}
From the relation $\{Q_{1},Q_{2}\}=0,$ it follows tha $\partial_{a}Q_{1}$
and $\partial_{a}Q_{2}$ are linearly dependent.

We can thus write:\foreignlanguage{english}{
\begin{equation}
\partial_{a}Q_{1}=\alpha(\sigma^{1},\sigma^{2})\partial_{a}Q_{2}
\end{equation}
for some coefficient function $\alpha(\sigma^{1},\sigma^{2})$}\footnote{%
The assumption that $\alpha$ is independent of $\sigma^{3}$ is based
on the premise that $Q_{1}$ and $Q_{2}$ are sufficiently smooth
functions of $\sigma^{3}$, with no singularities in that direction.
On the other hand, local irregularities or non-smooth behavior in
the $\sigma^{1}$ and $\sigma^{2}$ directions are allowed, as they
do not pose structural problems for the theory.%
}\foreignlanguage{english}{.}

Using this, we compute $K_{(\tau)}^{b}=\epsilon^{ab}\partial_{a}\partial_{\sigma^{3}}\tau(Q_{1},Q_{2})$
as:

\begin{equation}
\begin{aligned}\epsilon^{ab}\partial_{a}\partial_{\sigma^{3}}\tau(Q_{1},Q_{2})= & \epsilon^{ab}\partial_{\sigma^{3}}\tau(\partial_{a}Q_{1},Q_{2})+\epsilon^{ab}\partial_{\sigma^{3}}\tau(Q_{1},\partial_{a}Q_{2})\\
= & \epsilon^{ab}\alpha(\sigma^{1},\sigma^{2})\partial_{\sigma^{3}}\tau(\partial_{a}Q_{2},Q_{2})+\epsilon^{ab}\alpha^{-1}(\sigma^{1},\sigma^{2})\partial_{\sigma^{3}}\tau(Q_{1},\partial_{a}Q_{1}).
\end{aligned}
\end{equation}

Now, using the definition of $\tau(A,B)=\partial_{\sigma^{3}}A\cdot B-\partial_{\sigma^{3}}B$$\cdot A$,
we obtain:
\begin{equation}
\begin{aligned}= & \alpha(\sigma^{1},\sigma^{2})\epsilon^{ab}\partial_{\sigma^{3}}(\partial_{a}\partial_{\sigma^{3}}Q_{2}Q_{2}-\partial_{\sigma^{3}}Q_{2}\partial_{a}Q_{2})\\
 & +\alpha(\sigma^{1},\sigma^{2})^{-1}\epsilon^{ab}\partial_{\sigma^{3}}\left(\partial_{\sigma^{3}}Q_{1}\partial_{a}Q_{1}-\partial_{\sigma_{3}}\partial_{a}Q_{1}Q_{1}\right).
\end{aligned}
\end{equation}

Next, we consider the sum $K_{\Sigma1}^{b}+K_{\Sigma2}^{b}$. 
\begin{equation}
K_{\Sigma1}^{b}+K_{\Sigma2}^{b}=\partial_{\sigma^{3}}\left(Q_{1}\epsilon^{ab}\partial_{\sigma^{3}}\partial_{a}Q_{2}-Q_{2}\epsilon^{ab}\partial_{\sigma^{3}}\partial_{a}Q_{1}\right).
\end{equation}

Using again the linear dependence between $\partial_{a}Q_{1}$ and
$\partial_{a}Q_{2}$, we can express this as:

\begin{equation}
=\partial_{\sigma^{3}}\left(\alpha(\sigma^{1},\sigma^{2})^{-1}Q_{1}\epsilon^{ab}\partial_{\sigma^{3}}\partial_{a}Q_{1}-\alpha(\sigma^{1},\sigma^{2})Q_{2}\epsilon^{ab}\partial_{\sigma^{3}}\partial_{a}Q_{2}\right).
\end{equation}
Summarizing the results:

For $K_{(\tau)}^{b}$, we have:
\begin{equation}
\begin{aligned}K_{(\tau)}^{b}= & \alpha(\sigma^{1},\sigma^{2})\epsilon^{ab}\partial_{\sigma^{3}}(\partial_{a}\partial_{\sigma^{3}}Q_{2}Q_{2}-\partial_{\sigma^{3}}Q_{2}\partial_{a}Q_{2})\\
 & +\alpha(\sigma^{1},\sigma^{2})^{-1}\epsilon^{ab}\partial_{\sigma^{3}}\left(\partial_{\sigma^{3}}Q_{1}\partial_{a}Q_{1}-\partial_{\sigma_{3}}\partial_{a}Q_{1}Q_{1}\right).
\end{aligned}
\end{equation}
For $K_{\Sigma1}^{b}+K_{\Sigma2}^{b}$, we have:
\begin{equation}
K_{\Sigma1}^{b}+K_{\Sigma2}^{b}=\partial_{\sigma^{3}}\left(\alpha(\sigma^{1},\sigma^{2})^{-1}Q_{1}\epsilon^{ab}\partial_{\sigma^{3}}\partial_{a}Q_{1}-\alpha(\sigma^{1},\sigma^{2})Q_{2}\epsilon^{ab}\partial_{\sigma^{3}}\partial_{a}Q_{2}\right).
\end{equation}
Now, adding these two expressions and multiplying the entire result
by $\alpha(\sigma^{1},\sigma^{2})$, we obtain:

\begin{equation}
\begin{aligned}K_{(\tau)}^{b}+K_{\Sigma1}^{b}+K_{\Sigma2}^{b}= & -\alpha(\sigma^{1},\sigma^{2})^{2}\epsilon^{ab}\partial_{\sigma^{3}}(\partial_{\sigma^{3}}Q_{2}\partial_{a}Q_{2})+\epsilon^{ab}\partial_{\sigma^{3}}(\partial_{\sigma^{3}}Q_{1}\partial_{a}Q_{1})\\
= & 0
\end{aligned}
\end{equation}
Using once again the linear dependence relation 
\begin{equation}
\partial_{a}Q_{1}=\alpha(\sigma^{1},\sigma^{2})\partial_{a}Q_{2},
\end{equation}
we obtain:
\begin{equation}
\epsilon^{ab}\partial_{\sigma^{3}}\left(\left(\alpha(\sigma^{1},\sigma^{2})\partial_{\sigma^{3}}Q_{2}+\partial_{\sigma^{3}}Q_{1}\right)\partial_{a}Q_{1}\right)=0.
\end{equation}
Integrating this with respect to $\sigma^{3}$, we find:
\begin{equation}
\left(\alpha(\sigma^{1},\sigma^{2})\partial_{\sigma^{3}}Q_{2}+\partial_{\sigma^{3}}Q_{1}\right)\partial_{a}Q_{1}=C_{a}(\sigma^{1},\sigma^{2}),
\end{equation}
where $C_{a}(\sigma^{1},\sigma^{2})$ is an integration constant in
the $\sigma^{3}$ direction.

Solving for $\partial_{a}Q_{1}$, we obtain:
\begin{equation}
\partial_{a}Q_{1}=\frac{C_{a}(\sigma^{1},\sigma^{2})}{\left(\alpha(\sigma^{1},\sigma^{2})\partial_{\sigma^{3}}Q_{2}+\partial_{\sigma^{3}}Q_{1}\right)}.
\end{equation}
Now, assuming that the gauge parameters are smooth in the $\sigma^{3}$
direction, the denominator must be independent of $\sigma^{3},$and
we can write:
\begin{equation}
\partial_{a}Q_{1}=\lambda_{1,a}(\sigma^{1},\sigma^{2}).
\end{equation}
Similarly, we also have:
\begin{equation}
\partial_{a}Q_{2}=\lambda_{2,a}(\sigma^{1},\sigma^{2}).
\end{equation}

Substituting these conditions back into the original equation
\begin{equation}
\begin{aligned}K_{(\tau)}^{b}+ & K_{\Sigma1}^{b}+K_{\Sigma2}^{b}=\\
 & \epsilon^{ab}\partial_{a}\partial_{\sigma^{3}}\tau(Q_{1},Q_{2})+\partial_{\sigma^{3}}\left(Q_{1}\epsilon^{ab}\partial_{\sigma^{3}}\partial_{a}Q_{2}-Q_{2}\epsilon^{ab}\partial_{\sigma^{3}}\partial_{a}Q_{1}\right)=0,
\end{aligned}
\end{equation}
we find:
\begin{equation}
\partial_{a}\partial_{\sigma^{3}}\tau(Q_{1},Q_{2})=0.
\end{equation}
Integrating this with respect to $\sigma^{a}$, and assuming that
the integration constant vanishes due to appropriate boundary conditions,
we conclude:
\begin{equation}
\partial_{\sigma^{3}}\tau(Q_{1},Q_{2})=0.
\end{equation}

From the above analysis, we have derived all of the parameter constraints
for RVPD from the gauge restriction condition:
\begin{equation}
\frac{\partial\tau(Q_{1},Q_{2})}{\partial\sigma^{3}}=0,
\end{equation}

\begin{equation}
\{Q_{1},Q_{2}\}=0,
\end{equation}

\begin{equation}
\frac{\partial}{\partial\sigma^{3}}\frac{\partial Q_{1,2}}{\partial\sigma^{a}}=0.
\end{equation}

Thus, the necessary conditions have been successfully established.

\subsubsection*{B.4.1 Technical Summary of the Proof}

The discussion in this section can be summarized as follows:
\begin{enumerate}
\item By integrating the gauge restriction condition
\begin{equation}
C_{I}\frac{\partial X^{I}}{\partial\sigma^{3}}=\sigma^{3}
\end{equation}
along the $\sigma^{3}$ direction, we obtain
\begin{equation}
C_{I}X^{I}=\frac{1}{2}(\sigma^{3})^{2}+f(\sigma^{1},\sigma^{2}),
\end{equation}
where $f(\sigma^{1},\sigma^{2})$ is an integration constant.
\item Since $f(\sigma^{1},\sigma^{2})$ is arbitrary, we require that the
gauge restriction condition remain unchanged under any volume-preserving
deformation (VPD), regardless of the choice of $f$.
\item As a result, the VPD parameters $Q_{1}$ and $Q_{2}$ are subject
to constraints such as
\begin{equation}
\{Q_{1},Q_{2}\}=0,\ \partial_{\sigma^{3}}\tau(Q_{1},Q_{2})=0,
\end{equation}
which define the Restricted Volume-Preserving Deformation (RVPD).
\item In order to eliminate solutions of the form
\begin{equation}
\{Q_{1},Q_{2}\}=\frac{C(\sigma^{1},\sigma^{2})}{\sigma^{3}},
\end{equation}
we assume sufficient smoothness with respect to $\sigma^{3}$. These
lead to $C(\sigma^{1},\sigma^{2})=0$, ensuring that $\{Q_{1},Q_{2}\}=0.$
\item Altogether, in order to preserve the freedom introduced by the arbitrary
integration constant $f$, the VPD must be strongly restricted, resulting
in the RVPD structure.
\end{enumerate}
This chain of logic shows that any deformation preserving the gauge
restriction condition must take a highly restricted form. In particular,
it implies that reparametrizations in the $\sigma^{3}$ direction
are effectively prohibited.

This means that the gauge transformation parameters along $\sigma^{3}$
cannot vary freely, as they are constrained by conditions such as
$\{Q_{1},Q_{2}\}=0$, which restricts the extent to which reparametrizations
in the $\sigma^{3}$ direction can be performed.

\subsection*{B.5 Conceptual Summary and Interpretation}

With this, we have completed the proofs of both the sufficient and
necessary conditions.

By introducing the gauge restriction condition and defining the Restricted
Volume-Preserving Deformation (RVPD) accordingly, we are able to preserve
key structural properties of the theory---such as the composition
law of transformations and invariance---even after quantization via
matrix regularization.

This approach is fundamentally different from the conventional idea
of gauge fixing as the elimination of redundant degrees of freedom.

Instead, it should be understood as a structural constraint imposed
in order to preserve the algebraic consistency of the theory.

\bibliography{myTexJabRefLib}

\begin{thebibliography}{10}

\bibitem{Banks_1997}
T.~Banks, W.~Fischler, S.~H. Shenker, and L.~Susskind.
\newblock {M theory as a matrix model: A conjecture}.
\newblock {\em Physical Review D}, 55(8):5112--5128, April 1997.

\bibitem{de_Wit_1988}
B.~de~Wit, J.~Hoppe, and H.~Nicolai.
\newblock On the quantum mechanics of supermembranes.
\newblock {\em Nuclear Physics B}, 305(4):545--581, December 1988.

\bibitem{Fujikawa_1997}
Kazuo Fujikawa and Kazumi Okuyama.
\newblock On a {Lorentz} covariant matrix regularization of membrane theories.
\newblock {\em Physics Letters B}, 411(3-4):261--267, October 1997.

\bibitem{Awata_1998}
Hidetoshi Awata and Djordje Minic.
\newblock Comments on the problem of a covariant formulation of matrix theory.
\newblock {\em Journal of High Energy Physics}, 1998(04):006, April 1998.

\bibitem{Minic2000}
Djordje Minic.
\newblock Towards covariant {Matrix theory}.
\newblock {\em arXiv}, 2000.

\bibitem{Smolin_2000}
Lee Smolin.
\newblock {M theory as a matrix extension of Chern-Simons theory}.
\newblock {\em Nuclear Physics B}, 591(1-2):227--242, December 2000.

\bibitem{Yoneya_2016}
Tamiaki Yoneya.
\newblock {Covariantized matrix theory for D-particles}.
\newblock {\em Journal of High Energy Physics}, 2016(6):1--49, June 2016.

\bibitem{Ashwinkumar_2021}
Meer Ashwinkumar, Lennart Schmidt, and Meng-Chwan Tan.
\newblock {Matrix regularization of classical Nambu brackets and super
  p-branes}.
\newblock {\em Journal of High Energy Physics}, 2021(7), July 2021.

\bibitem{Awata_2001}
Hidetoshi Awata, Miao Li, Djordje Minic, and Tamiaki Yoneya.
\newblock {On the quantization of Nambu brackets}.
\newblock {\em Journal of High Energy Physics}, 2001(02):013--013, February
  2001.

\bibitem{Nambu_1973}
Yoichiro Nambu.
\newblock {Generalized Hamiltonian Dynamics}.
\newblock {\em Physical Review D}, 7(8):2405--2412, April 1973.

\bibitem{Takhtajan_1994}
Leon Takhtajan.
\newblock {On foundation of the generalized Nambu mechanics}.
\newblock {\em Communications in Mathematical Physics}, 160(2):295--315,
  February 1994.

\bibitem{Dito_1997}
G.~Dito, M.~Flato, D.~Sternheimer, and L.~Takhtajan.
\newblock {Deformation quantization and Nambu Mechanics}.
\newblock {\em Communications in Mathematical Physics}, 183(1):1--22, January
  1997.

\bibitem{Yoneya_2017}
Tamiaki Yoneya.
\newblock {Generalized Hamilton-Jacobi theory of Nambu mechanics}.
\newblock {\em Progress of Theoretical and Experimental Physics},
  2017(2):023A01, February 2017.

\bibitem{Saitou_2014}
M.~Saitou, K.~Bamba, and A.~Sugamoto.
\newblock {Hydrodynamics on non-commutative space: A step toward hydrodynamics
  of granular materials}.
\newblock {\em Progress of Theoretical and Experimental Physics},
  2014(10):103B03, October 2014.

\bibitem{Bagger_2007}
Jonathan Bagger and Neil Lambert.
\newblock {Modeling multiple M2-branes}.
\newblock {\em Physical Review D}, 75(4):045020, February 2007.

\bibitem{Gustavsson_2009}
Andreas Gustavsson.
\newblock {Algebraic structures on parallel M2 branes}.
\newblock {\em Nuclear Physics B}, 811(1-2):66--76, April 2009.

\bibitem{Aharony_2008}
Ofer Aharony, Oren Bergman, Daniel~Louis Jafferis, and Juan Maldacena.
\newblock {N = 6 superconformal Chern-Simons-matter theories, M2-branes and
  their gravity duals}.
\newblock {\em Journal of High Energy Physics}, 2008(10):091, October 2008.

\bibitem{Katagiri2022}
So~Katagiri.
\newblock {Quantization of Nambu Brackets from Operator Formalism in Classical
  Mechanics}.
\newblock {\em International Journal of Modern Physics A}, 38(18-19), July
  2022.

\bibitem{Nambu_1980}
Y.~Nambu.
\newblock {Hamilton-Jacobi formalism for strings}.
\newblock {\em Physics Letters B}, 92(3-4):327--330, May 1980.

\bibitem{Sugamoto_1983}
Akio Sugamoto.
\newblock Theory of membranes.
\newblock {\em Nuclear Physics B}, 215(3):381--406, February 1983.

\bibitem{Sakakibara_2000}
M.~Sakakibara.
\newblock {Remarks on a Deformation Quantization of the Canonical Nambu
  Bracket}.
\newblock {\em Progress of Theoretical Physics}, 104(5):1067--1071, November
  2000.

\bibitem{Matsuo_2001}
Yutaka Matsuo and Yuuichirou Shibusa.
\newblock Volume preserving diffeomorphism and noncommutative branes.
\newblock {\em Journal of High Energy Physics}, 2001(02):006, February 2001.

\bibitem{Ishibashi_1997}
Nobuyuki Ishibashi, Hikaru Kawai, Yoshihisa Kitazawa, and Asato Tsuchiya.
\newblock {A large-N reduced model as superstring}.
\newblock {\em Nuclear Physics B}, 498(1--2):467--491, August 1997.

\bibitem{katagiriSusy}
So~Katagiri.
\newblock Supersymmetric m2-brane matrix model with restricted
  volume-preserving deformations: Lorentz covariance and bps spectrum.

\end{thebibliography}

\end{document}